PLANETARY SCIENCE

# Krypton isotopes and noble gas abundances in the coma of comet 67P/Churyumov-Gerasimenko

Martin Rubin[1]*, Kathrin Altwegg[1,2], Hans Balsiger[1], Akiva Bar-Nun[3]†, Jean-Jacques Berthelier[4], Christelle Briois[5], Ursina Calmonte[1], Michael Combi[6], Johan De Keyser[7], Björn Fiethe[8], Stephen A. Fuselier[9,10], Sebastien Gasc[1], Tamas I. Gombosi[6], Kenneth C. Hansen[6], Ernest Kopp[1], Axel Korth[11], Diana Laufer[3], Léna Le Roy[1], Urs Mall[11], Bernard Marty[12], Olivier Mousis[13], Tobias Owen[14]†, Henri Rème[15,16], Thierry Sémon[1], Chia-Yu Tzou[1], Jack H. Waite[8], Peter Wurz[1,2]

The Rosetta Orbiter Spectrometer for Ion and Neutral Analysis mass spectrometer Double Focusing Mass Spectrometer on board the European Space Agency's Rosetta spacecraft detected the major isotopes of the noble gases argon, krypton, and xenon in the coma of comet 67P/Churyumov-Gerasimenko. Earlier, it was found that xenon exhibits an isotopic composition distinct from anywhere else in the solar system. However, argon isotopes, within error, were shown to be consistent with solar isotope abundances. This discrepancy suggested an additional exotic component of xenon in comet 67P/Churyumov-Gerasimenko. We show that krypton also exhibits an isotopic composition close to solar. Furthermore, we found the argon to krypton and the krypton to xenon ratios in the comet to be lower than solar, which is a necessity to postulate an addition of exotic xenon in the comet.

## INTRODUCTION

The European Space Agency's (ESA) Rosetta spacecraft accompanied comet 67P/Churyumov-Gerasimenko (hereafter 67P/C-G) over 2 years (1). In August 2014, the comet moved from a heliocentric distance of more than 3 astronomical units (AU) to perihelion in August 2015 at 1.24 AU and again away from the Sun. The Rosetta Orbiter Spectrometer for Ion and Neutral Analysis (ROSINA) monitored the abundances of volatiles in the coma of the comet in situ at the location of the spacecraft (2). The major components in the coma have been shown to be $H_2O$, $CO_2$, CO, and $O_2$ (3, 4), with strong variations depending on the location of the spacecraft with respect to the comet's heliocentric distance and solar illumination angle (5). As the comet approached the Sun, the activity increased and peaked some 2 to 3 weeks after perihelion passage (6). As the environment became increasingly dusty close to perihelion and the Rosetta star trackers encountered problems distinguishing stars from illuminated grains of cometary origin, the spacecraft was moved to larger cometocentric distances with lower dust abundances.

Volatile species measured in the coma have been released from the nucleus. Previous studies found that most of the ROSINA results could be explained without having to invoke significant contributions from distributed sources (6, 7). However, the changing illumination conditions and hence temperatures on the surface and the strong dichotomy of the northern and southern hemispheres of the nucleus, which hints at different processing of the surface (8, 9), make it difficult to derive ratios of the bulk volatile composition inside the nucleus. These relative abundances, however, are crucial to study the composition of the nascent solar system and the physical conditions under which it formed.

Early in the mission, Rosetta spent an extended amount of time close to the nucleus. In October 2014, at a 10-km distance from the nucleus, ROSINA detected $^{36}Ar$ and $^{38}Ar$, the two stable primordial noble gas isotopes of argon, respectively (10). The $^{36}Ar/^{38}Ar$ ratio was measured to be 5.4 ± 1.4, and this ratio is in agreement with the solar abundance ratio of 5.37 (11) within (large) errors. The ROSINA observations also revealed a remarkable correlation of argon with molecular nitrogen, $N_2$, of $^{36}Ar/N_2 = (9.1 ± 0.3) \times 10^{-3}$, while the ratio with respect to $H_2O$ showed a much larger variation of $^{36}Ar/H_2O = (0.1$ to $2.3) \times 10^{-5}$. These measurements were obtained far from perihelion at a distance of 3.1 AU. Late in the mission, in May 2016 when Rosetta was close to the nucleus of the comet, xenon was also detected. The seven major stable xenon isotopes ($^{128}Xe$, $^{129}Xe$, $^{130}Xe$, $^{131}Xe$, $^{132}Xe$, $^{134}Xe$, and $^{136}Xe$) found in the coma of 67P/C-G (12) showed an isotopic composition remarkably different from solar (12, 13). The heavy isotopes, $^{134}Xe$ and $^{136}Xe$, were depleted, while, on the other hand, $^{129}Xe$ showed enrichment with respect to solar and to chondritic values. Two possible causes were discussed (12): on the one hand, an r-process isotope deficiency relative to solar (or excess of s-process isotopes, which is equivalent) with an additional source for $^{129}Xe$ [possibly from the decay of now extinct $^{129}I$ with a half-life of 15.7 million years (Ma)] and on the other hand, a mass-dependent fractionation, which was then deemed improbable given the 67P/C-G xenon isotopic pattern. The accuracy of the measurements, however, did not allow for an unambiguous identification of the details of the underlying processes. Remarkably, 67P-C-G xenon fitted the $^{134}$Xe-$^{136}$Xe depletion well (relative to solar or chondritic) required to make the ancestor of terrestrial xenon (labeled Xe-U) and permitted to estimate the proportion of cometary xenon in the terrestrial atmosphere (*12*).

[1]Physikalisches Institut, University of Bern, Sidlerstrasse 5, CH-3012 Bern, Switzerland. [2]Center for Space and Habitability, University of Bern, Gesellschaftsstrasse 6, CH-3012 Bern, Switzerland. [3]Department of Geophysics, Tel Aviv University, Ramat-Aviv, Tel Aviv, Israel. [4]Laboratoire Atmosphères, Milieux, Observations Spatiales, Institut Pierre Simon Laplace, CNRS, Université Pierre et Marie Curie, 4 Place Jussieu, 75252 Paris Cedex 05, France. [5]Laboratoire de Physique et Chimie de l'Environnement et de l'Espace, UMR 6115 CNRS–Université d'Orléans, Orléans, France. [6]Department of Climate and Space Sciences and Engineering, University of Michigan, 2455 Hayward, Ann Arbor, MI 48109, USA. [7]Koninklijk Belgisch Instituut voor Ruimte-Aeronomie–Institut Royal Belge d'Aéronomie Spatiale, Ringlaan 3, B-1180 Brussels, Belgium. [8]Institute of Computer and Network Engineering, Technische Universität Braunschweig, Hans-Sommer-Straße 66, D-38106 Braunschweig, Germany. [9]Space Science Directorate, Southwest Research Institute, 6220 Culebra Road, San Antonio, TX 78228, USA. [10]University of Texas at San Antonio, San Antonio, TX 78249, USA. [11]Max-Planck-Institut für Sonnensystemforschung, Justus-von-Liebig-Weg 3, 37077 Göttingen, Germany. [12]Centre de Recherches Pétrographiques et Géochimiques, CNRS, Université de Lorraine, 15 rue Notre Dame des Pauvres, BP 20, 54501 Vandoeuvre lès Nancy, France. [13]Laboratoire d'Astrophysique de Marseille, CNRS, Aix-Marseille Université, 13388 Marseille, France. [14]Institute for Astronomy, University of Hawaii, Honolulu, HI 96822, USA. [15]Institut de Recherche en Astrophysique et Planétologie, CNRS, Université Paul Sabatier, Observatoire Midi-Pyrénées, 9 Avenue du Colonel Roche, 31028 Toulouse Cedex 4, France. [16]Centre National d'Études Spatiales, 2 Place Maurice Quentin, 75001 Paris, France.
*Corresponding author. Email: martin.rubin@space.unibe.ch
†Deceased.

During the same time period late in the Rosetta mission, the isotopes of krypton, namely, $^{80}$K, $^{82}$Kr, $^{83}$Kr, $^{84}$Kr, and $^{86}$Kr, have also been detected. Here, we report on their isotopic ratios and the corresponding relative abundances of the noble gases argon, krypton, and xenon (neon was not detected).

## RESULTS
### ROSINA Double Focusing Mass Spectrometer
ROSINA consists of the Double Focusing Mass Spectrometer (DFMS), the Reflectron-type Time-Of-Flight (RTOF) mass spectrometer, and the COmet Pressure Sensor (COPS). The two mass spectrometers complement each other with high mass resolution (DFMS) and high time resolution (RTOF). Here, we focus on neutral composition measurements performed by ROSINA DFMS. In neutral mode, particles are first ionized through electron impact and then separated on the basis of their mass-to-charge ratio in an electric and then a magnetic field before hitting the detector. All errors given in this paper are 1-σ. More details about DFMS and the data analysis can be found in Materials and Methods.

### DFMS observations in May 2016
In the late portion of the mission, Rosetta spent more time close to the nucleus, and ROSINA detected the noble gas xenon (*12*). For this purpose, we investigated ROSINA DFMS data obtained between 14 and 31 May 2016 between 10 and 7 km from the nucleus' center of mass. The observations were performed 2 months after the outbound equinox and mostly above the southern hemisphere that remained more active despite the subsolar latitude moving to the northern hemisphere. As the spacecraft was slowly approaching the comet, the DFMS detector count rates slightly increased. Three overlapping time periods were analyzed, as follows: (i) 14 to 31 May 2016, (ii) 20 to 31 May 2016, and (iii) 26 to 31 May 2016. Period (i) contained approximately 500 individual mass spectra but had the lowest average count rates, period (ii) was an average of approximately 260 spectra with medium average count rates and was also contained in period (i), and period (iii) was based on approximately 70 spectra with the highest average count rates and was also part of both periods (i) and (ii). Similar results were obtained for all three periods, and the analysis was focused on period (ii), which was a compromise of high enough average count rates to investigate the less abundant $^{128}$Xe and $^{130}$Xe and the number of spectra used in the averaging process.

During the same 2.5 weeks in May 2016, when Rosetta was inside of 15 km from the comet center, ROSINA also detected the isotopes of the noble gas krypton ($^{80}$Kr, $^{82}$Kr, $^{83}$Kr, $^{84}$Kr, and $^{86}$Kr). We performed the data analysis for the same three periods as for xenon. However, on the basis of the higher instrument sensitivity for krypton versus xenon and the resulting higher average count rates, we focused on period (i), which contains the full data set of more than 500 spectra per mass line and includes periods (ii) and (iii). The data analysis was the same as for xenon (*12*), and details are in Materials and Methods. Figure S1 shows the averages of period (i) of the 500 individual spectra for each isotope ($^{82}$Kr, $^{83}$Kr, $^{84}$Kr, and $^{86}$Kr), and Fig. 1 shows the isotopic ratios normalized to $^{84}$Kr and to the solar wind (SW) composition (discussion follows later). The mass spectra in fig. S1 also show fragments of hexane, $C_6H_{14}$, and higher-mass organic species, which shall be discussed elsewhere.

The $^{80}$Kr mass line has a strong interference with carbon disulfide with two heavy sulfur isotopes, $C^{34}S_2$. The resulting large error bars for $^{80}$Kr did not allow a meaningful contribution to our analysis, and thus, we also omitted the corresponding spectra from the figure. $^{78}$Kr is unfortunately fully hidden below carbon disulfide containing one heavy sulfur isotope, $C^{34}S^{32}S$, and therefore excluded from our analysis. Table S1 shows the corresponding results for the three periods (i) to (iii). The counts represent the sum of counts under the peak (sum of the red points in fig. S1) averaged per isotope with the statistical count, offset, and fit errors indicated below.

The noble gas measurements in May 2016 were performed above the southern hemisphere during ever-closer orbits in the terminator plane as the comet rotated once per approximately 12 hours. Still, at these distances, the whole comet was in the field of view of DFMS, and gas may have come from any location on the nucleus facing Rosetta. During the 2.5 weeks of observations, the heliocentric distance of the comet increased only slightly. These small changes and the fact that there was no spacecraft slewing (that could change the background) thus provided stable observation conditions throughout the period chosen. Figure S2 shows Rosetta's position with respect to the comet for the time of observation and the three investigated time intervals.

### Relative noble gas abundances from May 2016
In all three periods, the noble gas argon has been also detected. By combining these observations, the relative abundances of the $^{36}$Ar, $^{84}$Kr, and $^{132}$Xe noble gas isotopes are obtained. Figure 2A shows the abundances of $^{84}$Kr versus $^{36}$Ar. Because of the low count rates for krypton, daily averages are plotted, with the day of the month indicated next to the measurement. Figure 2 shows a reasonable correlation between $^{36}$Ar and $^{84}$Kr. The interference of $^{36}$Ar with $H^{35}Cl$ and $H_2^{34}S$ (*10*, *14*, *15*) leads to larger error bars compared to the well-separated $^{84}$Kr peak. The slope of $^{84}$Kr/$^{36}$Ar = 0.089 ± 0.021 indicates their relative abundance. Figure 2B shows the same features but for $^{132}$Xe as a function of $^{84}$Kr, which exhibits a slope of $^{132}$Xe/$^{84}$Kr = 0.269 ± 0.064. In addition to errors from counting statistics, the errors of the slopes contain 16% uncertainty of the measured sensitivities of argon, krypton, and xenon obtained in the laboratory calibration campaign (more details in the Supplementary Materials). Not all daily averages contain the same number of spectra, and for 30 May 2016, there were only two spectra, which was not enough to derive a xenon abundance. On 28 and 29 May 2016, no measurements were performed because of an outage of the Rosetta star trackers caused by excessive dust and the ensuing safe mode of the spacecraft. These results correspond to the ratio of the local densities of $^{36}$Ar/$^{132}$Xe = 41.7 ± 10.1, $^{36}$Ar/$^{84}$Kr = 11.2 ± 2.6, and $^{84}$Kr/$^{132}$Xe = 3.7 ± 0.9.

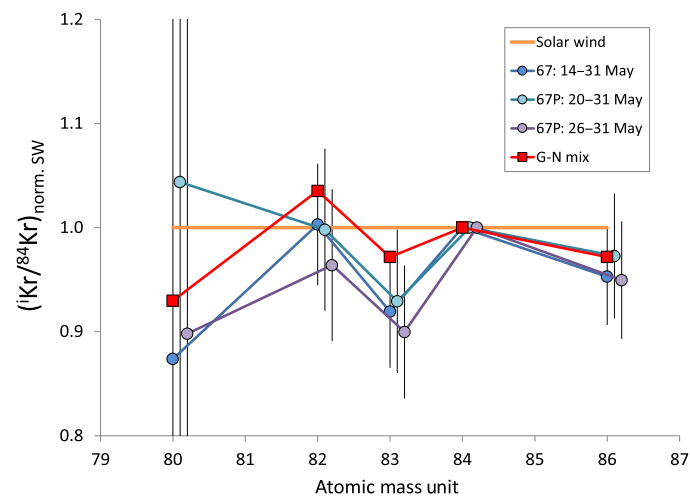

**Fig. 1. Isotopic composition of 67P/C-G krypton, normalized to $^{84}$Kr and the SW composition [from (*18*)].** 67P/C-G errors reflect 1-σ SEM and calibration uncertainties for the corresponding averaging periods. In this format, SW-Kr is represented by the horizontal orange line. $^{83}$Kr appears to be slightly depleted relative to solar. The red line represents a mix of different nucleosynthetic components [the so-called G-Kr and N-Kr components; (*11*, *20*)]. For the G-Kr composition, we consider the weak s-process composition having low $^{86}$Kr/$^{84}$Kr ratios (*20*). The best fit was obtained for a proportion of 5% G-Kr in cometary krypton.

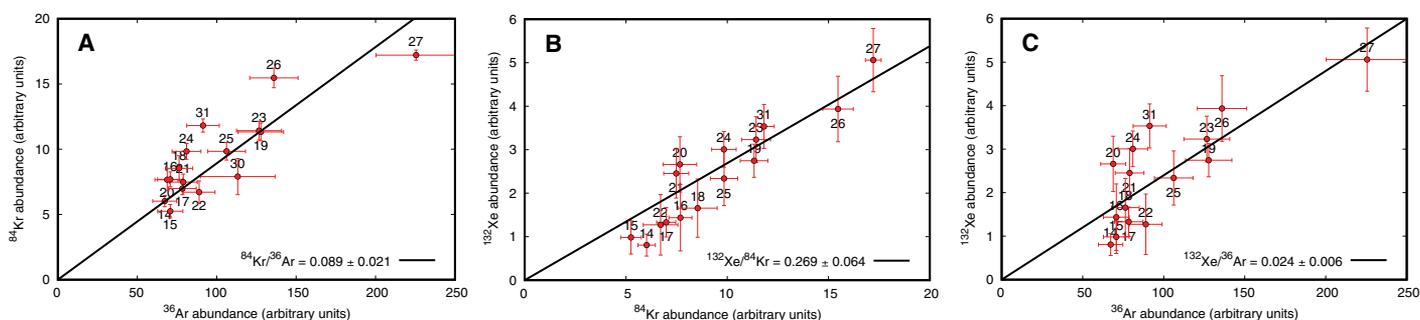

**Fig. 2. Relative abundances of the major isotopes of argon, krypton, and xenon.** (**A**) $^{84}$Kr versus $^{36}$Ar relative abundances obtained from daily averages with SEM error bars for the 14 to 31 May 2016 period ($R^2$ = 0.67). (**B**) $^{132}$Xe versus $^{84}$Kr. (**C**) $^{132}$Xe versus $^{36}$Ar. The number associated with each point indicates the day ($R^2$ = 0.83). The error of the daily averages includes the statistical errors, whereas for the slopes, a 16% calibration uncertainty for each element is included. A correlation coefficient of 0.76 was found between the $^{132}$Xe/$^{36}$Ar and $^{84}$Kr/$^{36}$Ar slopes (A and C).

However, these slopes might not reflect the actual relative noble gas production rates of the comet as there is a local enhancement of the density of heavier species due to the lower outgassing velocity. Equal outgassing velocities can only be expected in the dense, collisionally coupled coma. At the location of the spacecraft, the coma of 67P/C-G was always very thin, with very long collisional mean free paths. Numerical simulations based on the Direct Simulation Monte Carlo technique (6, 16) have been carried out through a range of heliocentric distances showing that, beyond 3 AU, the individual species collisionally decouple directly at the surface. We therefore applied a correction factor assuming a $1/\sqrt{\text{mass}}$ dependence of the outgassing velocity and the corresponding production rate ratio from the nucleus. This assumption implies the same temperatures of the species in the gas phase after passing the surface dust layer through thermal accommodation by collisions and the same (subsurface) source regions for the different noble gases (and N$_2$), which is compatible with the observed correlations shown in Fig. 2. We thus obtain for the relative production rate at the surface of the comet $^{36}$Ar/$^{132}$Xe = 79.8 ± 19.3, $^{36}$Ar/$^{84}$Kr = 17.1 ± 3.9, and $^{84}$Kr/$^{132}$Xe = 4.7 ± 1.1. Table 1 provides a collection of the results with and without applied velocity correction factors. On the other hand, to be able to compare with published argon (10) and xenon (12) data, we have omitted the velocity correction in relative abundances of the detected krypton isotopes that amounts to, for example, 1.2% in the case of $^{86}$Kr/$^{84}$Kr, which is well within the reported 1-σ error of 4.9% (Fig. 1).

### Noble gas bulk abundances in the nucleus of 67P/C-G

We undertook an attempt to derive noble gas bulk abundances relative to water inside the nucleus. Briefly, through molecular nitrogen, N$_2$, we were able to combine measurements obtained late in the mission when Rosetta was close enough to the comet to measure Ar/N$_2$, Kr/Ar, and Xe/Ar with measurements of N$_2$/H$_2$O close to perihelion, which better reflect bulk abundances inside the nucleus but were obtained too far from the comet to measure noble gases. The results listed in Table 2 were limited to the southern hemisphere, which was responsible for the major fraction of the total outgassing during that time (17) and where fresh material is therefore exposed during each pericenter passage (9). Any species and temperature dependence in the sublimation process would have to be considered on top of the numbers presented in this paper (apart from the velocity correction).

**Table 1. Relative $^{36}$Ar, $^{84}$Kr, and $^{132}$Xe abundances for the 14 to 31 May 2016 period based on approximately 500 spectra.** The second column gives the density ratios measured at the location of Rosetta, which are equivalent to the comet's production rate, assuming equal outgassing velocities applicable in a collisional coma. The last column (underlined) represents our best estimate for the bulk abundance where the ratio $X_1/X_2$ is multiplied by $\sqrt{m_2/m_1}$ to account for different outgassing velocities based on a thermal expansion approximation. Errors reflect 1-σ SEM and calibration uncertainties.

| Isotope | Density ratio at Rosetta | $\sqrt{m_2/m_1}$ | Production rate ratio of 67P/C-G |
|---|---|---|---|
| $^{36}$Ar/$^{132}$Xe | 41.7 ± 10.1 | 1.915 | 79.8 ± 19.3 |
| $^{36}$Ar/$^{84}$Kr | 11.2 ± 2.6 | 1.527 | 17.1 ± 3.9 |
| $^{84}$Kr/$^{132}$Xe | 3.7 ± 0.9 | 1.254 | 4.7 ± 1.1 |

The first assumption is that noble gases are released in proportion to their relative abundances in the comet's nucleus. Relative abundances of $^{36}$Ar, $^{84}$Kr, and $^{132}$Xe are presented in the previous section. Furthermore, we derived the fraction of $^{36}$Ar isotopes with respect to the total amount of argon of 0.844 ± 0.034 (10). The same can be obtained for $^{84}$Kr/Kr = 0.582 ± 0.010 (this work) and $^{132}$Xe/Xe = 0.253 ± 0.012 (12), respectively (see also table S2).

Given the increased activity and hence the dusty near-nucleus environment, the Rosetta spacecraft was moved to larger cometocentric distances. Therefore, ROSINA could not detect the noble gases during the most active period of the comet. Inspired by earlier observations (10), we again investigated molecular nitrogen, N$_2$, which was abundant enough to be observed throughout the mission. While measured above the southern hemisphere this time, N$_2$ and $^{36}$Ar still show a strong correlation. Figure 3B shows the correlation between N$_2$ and $^{36}$Ar for period (i) and 2 months earlier during 9 to 21 March 2016 when the spacecraft was sufficiently close to the nucleus for the detection of argon. Figure 3C shows $N_2$ (and hence $^{36}Ar$) versus $CO_2$ and $^{36}Ar$ versus $H_2O$ (Fig. 3A). Both sets of measurements from May 2016 and March 2016 were obtained at different cometocentric distances, and hence, the local densities of $^{36}Ar$ at the spacecraft were higher for the period in May despite the larger heliocentric distance. The illumination of the nucleus also changed as the outbound equinox in March 2016 marked the start of summer in the northern hemisphere. The water production again followed the subsolar latitude to the northern hemisphere (*17*), while noble gases were still more dominant above the southern hemisphere. For these reasons, a marked change in the $^{36}Ar/H_2O$ ratio can be observed, while $^{36}Ar/N_2$ remains well correlated. The slope of $^{36}Ar/N_2$ of ~$6.2 \times 10^{-3}$ is somewhat less steep compared to the $9.1 \times 10^{-3}$ measured early in the mission (*10*) above the northern hemisphere, which showed lower relative abundances of highly volatile species (*5*), possibly due to the redeposition of wet dust (*8*). We will therefore base our results on the relative abundance of $^{36}Ar/N_2$ of ~$6.2 \times 10^{-3}$ measured above the freshly processed southern hemisphere and assume that $Ar/N_2$ remains more or less unchanged throughout the cometary orbit.

Table 2. **Best-estimate bulk abundances of noble gases in comet 67P/C-G in the second column.** The third column gives the correction factor applied to the values from the second column in table S3 based on the expected outgassing velocities. We assume that, close to perihelion, $N_2$ and $H_2O$ share the same velocity due to collisional coupling and apply no correction. Hence, for the May 2016 period, the reference mass for the $1/\sqrt{m}$ velocity correction becomes that of molecular nitrogen ($m_1 = 28$ u) instead of water (18 u). Errors reflect 1-σ SEM and calibration uncertainties.

| Species | Production rate ratio of 67P/C-G | $\sqrt{m_2/m_1}$ | Conditions/notes |
|---|---|---|---|
| $N_2/H_2O$ | $(8.9 \pm 2.4) \times 10^{-4}$ | (1.000) | Collisional, measured near perihelion |
| $^{36}Ar/N_2$ | $(5.5 \pm 1.5) \times 10^{-3}$ | 0.882 | Rarefied, measured beyond 3 AU |
| $^{36}A/H_2O$ | $(4.9 \pm 1.9) \times 10^{-6}$ | 0.882 | Combination of rarefied and collisional |
| $Ar/H_2O$ | $(5.8 \pm 2.2) \times 10^{-6}$ | 0.882·(1.000) | Combination of rarefied and collisional |
| $Kr/H_2O$ | $(4.9 \pm 2.2) \times 10^{-7}$ | 0.577 | Combination of rarefied and collisional |
| $Xe/H_2O$ | $(2.4 \pm 1.1) \times 10^{-7}$ | 0.461 | Combination of rarefied and collisional |
| $Ne/H_2O$ | $< 5 \times 10^{-8}$ | 1.128 | Upper limit |

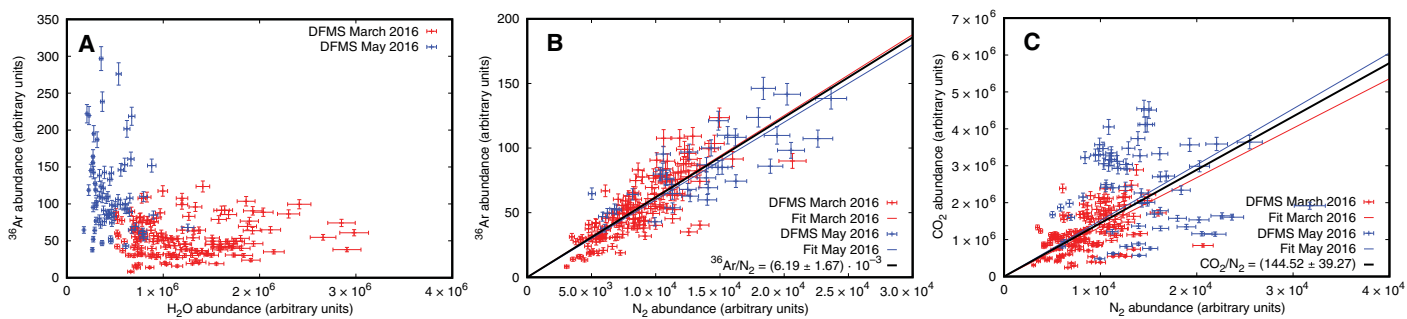

**Fig. 3. Relative abundances of $^{36}Ar$, $N_2$, and $CO_2$.** $^{36}Ar$ to $H_2O$ abundances for 9 to 21 March 2016 (red) and 14 to 31 May 2016 (blue) in (**A**), $^{36}Ar$ to $N_2$ abundances in (**B**), and $N_2$ to $CO_2$ in (**C**). The individual measurements show statistical errors. The error of the slope for the combined measurements (black) includes the statistical error, 18% calibration uncertainty (sensitivity and fragmentation pattern for each species), and 20% gain error ($R^2 = 0.70$ for $^{36}Ar/N_2$ and $R^2 = 0.12$ for $CO_2/N_2$).

In the next step, the period from May 2015 to early June 2015 was analyzed to estimate bulk abundances in the comet (*14*). Later on, around perihelion in August 2015, the coma was much more dynamic with frequent outbursts, which can potentially lead to an overestimation of the highly volatile species (such as $N_2$) with respect to their bulk abundance. Furthermore, a dedicated Philae lander search required a position above the lesser active northern hemisphere. After perihelion, Rosetta went to even larger distances for an excursion to the bow shock to study the interaction of the comet with the SW. Especially in May 2015, Rosetta not only flew over the active southern hemisphere at negative latitudes but also spent time on the dayside of the comet (phase angles down to 60°) for which we derived a bulk abundance ratio of $N_2/H_2O = (8.9 \pm 2.4) \times 10^{-4}$ (compare Table 2). Details on Rosetta's trajectory before perihelion for May 2015 to early June 2015 can be found in fig. S3. All measurements are restricted to the dayside of the comet (phase angle < 85°) above the southern summer hemisphere (subspacecraft latitude < 0°). Out of this selection, two subsets were analyzed: first, the period of 10 to 31 May 2015, and second, the full month of May and early June (period denoted by the two gray bars in fig. S3). Both periods resulted in very similar $N_2/H_2O$ ratios of ~$(8.9 \pm 2.4) \times 10^{-4}$, which we then adopted for the bulk abundance of $N_2$. Figure S4 shows the measured $N_2/H_2O$ abundance ratios and the corresponding fits.

Connecting these observations (for example, $^{36}Ar/Ar$, $^{36}Ar/N_2$, and $N_2/H_2O$) yields an estimation of the bulk abundances of the noble gas argon in comet 67P/C-G with respect to water. Then, including $^{84}Kr/Kr$, $^{132}Xe/Xe$, $^{36}Ar/^{84}Kr$, and $^{36}Ar/^{132}Xe$, we obtained the $Kr/H_2O$ and $Xe/H_2O$ ratios, respectively. The results are summarized in table S3, again derived directly from the local densities at Rosetta and corrected for different outgassing velocities. The former is applicable for a collisionally coupled coma, while the latter is applicable for a rarefied coma. Although the coma is no longer collisional at a few 100 km, where Rosetta was located in May 2015 and June 2015, it is likely that $N_2$ and $H_2O$ have similar outgassing velocities due to collisional coupling within the first few kilometers from the nucleus (*16*). Therefore, our best estimate in Table 2 assumes collisionally coupled $N_2$ and $H_2O$ close to the nucleus near perihelion and decoupled velocities for all measurements obtained in the May 2016 period beyond 3 AU.

Throughout the reported measurement period in May 2016, we also looked for the noble gas neon. The lack of a signal of $^{20}$Ne on mass/charge 20 u/e and of $^{22}$Ne on 22 u/e allows us to derive an upper limit for the neon relative abundance with respect to $H_2O$ of $^{21,22}$Ne/$H_2O \leq 5 \times 10^{-8}$. The rather low sensitivity of DFMS for neon and thus high upper limit are a consequence of its small cross section for electron impact ionization.

## DISCUSSION
### Constraints on the origin of cometary ice from the noble gas isotopic composition

The argon isotopic composition of 67P/C-G's coma [$^{36}$Ar/$^{38}$Ar = 5.4 ± 1.4; (*10*)] is not precise enough to distinguish between most solar system reservoirs including meteoritic [the so-called Q component, $^{36}$Ar/$^{38}$Ar = 5.34; see (*11*) and references therein] and solar [$^{36}$Ar/$^{38}$Ar = 5.37 (*11*)]. Given the error on the 67P/C-G measurement, the argon isotopic composition cannot be distinguished from nucleosynthetic components identified in meteoritic presolar grains, such as P3-Ar (5.26), N-Ar (5.87), or HL-Ar (4.41) [see (*11*) and references therein]. The isotopic composition of 67P/C-G–Xe is, however, not consistent with a solar-like origin, being depleted in the heavy xenon isotopes ($^{134}$Xe and $^{136}$Xe) and being rich in $^{129}$Xe relative to solar. This composition has been attributed to a nucleosynthetic mix different from that of the bulk solar system as opposed to be the result of isotopic mass fractionation, suggesting that cometary ice contains interstellar material (*12*). Thus, the isotopic composition of krypton may further constrain the origin of cometary material.

Figure 1 represents the krypton isotopic composition recorded in the coma of comet 67P/C-G in May 2016, normalized to $^{84}$Kr and to the SW composition (*18*). Within errors, the 67P/C-G-Kr composition is close to that of the SW, with the possible exception of a deficit in $^{83}$Kr. This similarity contrasts with the case of xenon in 67P/C-G, whose isotopic composition is clearly different from solar (*12*). The large excess in $^{129}$Xe was tentatively attributed to the decay of extinct $^{129}$I ($T_{1/2}$ = 16 Ma), provided that the initial iodine abundance in the parent reservoir of xenon was orders of magnitude higher than that present in meteorites at the beginning of solar system formation, thus implying a presolar origin for the parent xenon reservoir. The deficits of $^{134}$Xe and $^{136}$Xe could not originate from nuclear processes other than nucleosynthetic ones, and it was suggested that these deficits resulted from a mix of presolar components different from that of the bulk solar system (*12*). Adopting the nucleosynthetic end-members from correlations in presolar materials (*19*), Marty *et al.* argued (*12*) that the 67P/C-G xenon composition could be reproduced by mixing up an s-process xenon component with two r-process xenon components identified earlier (*19*).

In principle, a similar mix should be able to reproduce the krypton isotopic composition of the comet. Unfortunately, such an approach is not applicable to krypton because there are no r-process–only isotopes for krypton (*20*), and therefore, the identification of pure r-composition(s) cannot be achieved. Thus, the mixing approach done for xenon cannot be applied directly to krypton. The lack of r-process–only krypton isotopes may also explain why the krypton isotopic composition is less variable than that of xenon among solar system reservoirs and objects. Another complication is that the production rates of s-process krypton isotopes in asymptotic giant branch (AGB) stars can vary depending on the stellar regime. This is because $^{86}$Kr (and $^{80}$Kr) is affected by branching in the s-process, contrary to $^{82,83,84}$Kr isotopes. The abundances of $^{86}$Kr and $^{80}$Kr have been found to be variable among SiC grains (*20*), in proportions that are consistent with theoretical models [for example, (*21*)].

It has been observed (*20*) that the krypton isotopic ratios of SiC grains define good mixing correlations between two nucleosynthetic end-members identified as an s-process component labeled G-Kr and a "normal" component labeled N-Kr. The normal composition integrated the contribution of several nucleosynthetic sources and could represent the nucleosynthetic ancestor of solar system krypton. Later on, a qualitatively similar scenario was proposed (*22*) to derive the xenon and krypton isotopic compositions of meteorites. An exotic, weak s-process–rich material was added before solar system formation to an ancestral presolar component (labeled "P3") resembling the solar composition. We assume that the krypton isotopic composition of 67P/C-G is the result of the contribution of an exotic component, s-process (G) krypton to normal (N) krypton (*20*), as observed in SiC grains [end-member data summarized in (*11*)]. Based on this assumption, the 67P/C-G krypton composition, including the slight deficit in $^{83}$Kr, can be reproduced within errors by adding ~5% of the G (s-process) component to the N component (Fig. 1). Taking SW krypton instead of N-Kr would give comparable results because N-Kr is isotopically close to SW-Kr. The data are best fitted by assuming a weak s-process component having a low $^{86}$Kr/$^{84}$Kr ratio corresponding to low neutron flux in AGB envelopes (Fig. 1).

If this scenario is correct, then the contribution of s-process material before solar system formation must have also affected the xenon isotopic composition by increasing the s-process contribution relative to the r-process one, which is effectively what is required for 67P/C-G xenon (*12*). In the Supplementary Materials, we have modeled the compositions of krypton and xenon by assuming a similar mixing scenario for both noble gases. To a normal composition, represented by N-Kr and N-Xe [(*11*) for end-member compositions] an exotic component (X-Kr and X-Xe, rich in s-process isotopes) is added. For X-Kr, it is not possible to define the r-process contribution for the reason given above, and we use G-Kr (*20, 22*). For xenon, we use a similar mix of s-process and r-process components reproducing the cometary xenon composition (*12*). The proportions of the mix that best fit xenon data are 35% s-process xenon and 65% r-process xenon (the latter being the mean of the two r-process end-members (*19*). For the s-process xenon component, taking either the composition identified by (*19*) or the G-Xe composition gives similar results. For the sake of consistency with krypton, we select G-Xe. The best fit for the 67P/C-G composition is obtained for X-Xe = 80% and N-Xe = 20%.

The much higher proportion of the excess component added to xenon (80%) compared to that added to krypton (5%) implies that the (Xe/Kr)$_X$ ratio of the exotic component has to be higher than the (Xe/Kr)$_N$ ratio of the normal (N) component. For the sake of illustration, we assume below that the N-(Xe/Kr) ratio is similar to the solar ratio. In agreement with this requirement, we note that the Xe/Kr ratio of the comet is 4.7 ± 1.8 times the solar ratio [from data in Table 1 and (13)]. We compute that the $(Xe/Kr)_X$ ratio of the exotic component is ~80 times the solar ratio for a proportion of 5% G-Kr (mixing equations and calculations are given in the Supplementary Materials). We can also compute what would be the corresponding fraction of exotic xenon in cometary xenon, which gives 80%, the remaining 20% being N-Xe. This fraction is in agreement with a fraction of 80% exotic xenon required independently to fit best the xenon isotopic data (fig. S5).

The mixing model proposed above is able to provide a satisfactory solution for explaining the contrasted isotopic compositions of krypton and xenon. This scenario is, however, not without problems since it requires the exotic, s–process–rich component in the comet to be significantly enriched in xenon relative to krypton compared to the solar abundance. Studies of presolar material have shown that xenon is commonly enriched relative to krypton in the s-process end-members, often referred as chemical fractionation (20). These enrichments could be related to differences in the behaviors of xenon and krypton during expulsion from stellar envelopes, such as preferential ionization of xenon relative to krypton, and/or selective implantation of xenon into dust. It is unclear whether such chemical fractionation could have played a role in enriching icy grains with AGB-derived noble gases. Selective trapping or retention of xenon during ice-related processes could have also played a role, and experiments involving noble gas trapping and desorption in ice under conditions relevant to the outer solar system are highly needed.

### Noble gas abundance of solar system reservoirs
In Fig. 4, the relative abundances of 67P/C-G noble gases (Table 2) are represented in an $^{84}Kr/^{36}Ar$ versus $^{132}Xe/^{36}Ar$ format, together with solar, Earth, Mars (11), and chondritic [Centre de Recherches Pétrographiques et Géochimiques (CRPG), Nancy compilation of Carbonaceous Orgueil (CI) and Carbonaceous Murchison (CM) data] noble gas compositions including upper limits for Titan (23). 67P/C-G is clearly depleted in argon relative to solar, which can be accounted for by an ice trapping temperature that would not permit complete retention of argon or additional contributions of xenon (and krypton). Noble gas trapping in amorphous ice as a function of temperature is represented by the thin blue line and blue bars given by (24) based on data from (25). Taken at face value, 67P/C-G noble gas relative abundances would require relatively high trapping temperature (≥70 K), which is not consistent with other 67P/C-G coma measurements. For instance, the $CO/N_2$ ratio requires temperatures ≤50 K (26).

Relative abundances of volatile species in the coma of comet 67P/C-G are the result of their trapping efficiencies in the cometesimals, as well as release and diffusion throughout the porosity of cometary ices. Comet 67P/C-G belongs to the group of Jupiter-Family Comets (JFCs), which most likely spent most of their life in the scattered disk. These objects do not easily become JFCs but are thought to have undergone a series of gravitational scattering processes, which leave them for several million years as Centaurs at intermediate distances from the Sun (27), while comets from the Oort Cloud (OCC) have a very different dynamical history, and their trajectories differ significantly from those of JFCs.

The thermal evolution of comet 67P/C-G assuming 10 Ma in a Centaur orbit at 7 AU has been simulated (28). These simulations yield inhomogeneous temperatures up to 80 to 90 K inside the nucleus at a depth of at least a kilometer. In combination with the amorphous to crystalline transition or any kind of destabilization of ices, this leads to nonhomogeneous subsurface composition as an evolutionary process, even before the comet's first apparition in the inner solar system.

At these temperatures and over a time span of several million years, species of high volatility may partially diffuse from the comet and be lost (29). As a consequence, the abundance of highly volatile species may be altered in JFCs and different from OCCs, provided that the bulk outgassing occurs from an altered layer. The CO production rate of 67P/C-G around perihelion is ~1% compared to water. This is toward the low end of CO production rates compared to other comets, where production rates extend up to 25% (30). Comets are depleted in nitrogen by at least a factor of 5. This was explained by not incorporating the full complement of $N_2$ or by loss of $N_2$ during the lifetime of the comet (31). The low abundance of the highly volatile CO and $N_2$ (26) could have the same origin, namely, selective diffusion in amorphous ice. This would then also have led to loss of argon and, if initially present at all, neon.

The erosion of 67P/C-G is estimated to be a few tens of meters per pericenter passage (9). The comet has been in its current orbit since a close encounter with Jupiter in 1959, but even before that, it may have been inside of 5 AU for several centuries (32), and thus, several hundreds of meters of surface erosion could have significantly altered the size and shape of the comet. It may therefore have lost most of the layer affected by the heat wave during the Centaur stage, but this is difficult to assess.

Another possibility for the apparent mismatch of our results compared to amorphous ice laboratory measurements is the specific set of conditions (for example, gas composition and pressure) that are not representative of conditions in nature. The temperature of release of volatile species, in particular from amorphous ice, can be significantly different from their trapping temperature and depends on the ice in which they are embedded (33). Hence, Kouchi and Yamamoto (34) also investigated the trapping and release of gases in mixtures better representing the volatile species in a comet. In particular, the non-negligible abundance of $CO_2$ influences the trapping and subsequent release of volatile species from amorphous ices. Parts of the $CO_2$ are embedded in the water ice, while also patches of $CO_2$ ice form. With a mixture of 65:10:10:15 of $H_2O:CO_2:CH_4:CO$, the very volatile species, CO and $CH_4$, trap much better and are released in a complicated pattern compared to mixtures with pure water ice only. The same behavior can be observed for argon and $N_2$ in mixtures of $H_2O$ and $CO_2$, which raises the question on the (relative) trapping behavior of the other noble gases neon, krypton, and xenon (35). From mixtures of $H_2O$:Ne at different temperatures between 15 and 35 K, the Ne/$H_2O$ ratio drops sharply already at 35 K to the value of $10^{-4}$ (36). The gas is released from the amorphous ice during the transformations to crystalline ice from 130 to 160 K. The estimated upper limit of $5 \times 10^{-8}$ would rise the cometary ice formation temperature to 40 K. However, the release of krypton and xenon in 67P/C-G was detected from the southern, winter hemisphere very late in the mission. At this time in the mission, water sublimation from the south was more or less shut off, while $CO_2$ was still abundant (17). The abundant $CO_2$ relative to water suggests that these noble gases are predominantly trapped in the $CO_2$ fraction of the ice. This may also explain the poor correlation of argon with $H_2O$ (10). The interpretation of our results requires more laboratory experiments, where trapping in more comet-like mixtures, with $H_2O$ and at least $CO_2$, is investigated.

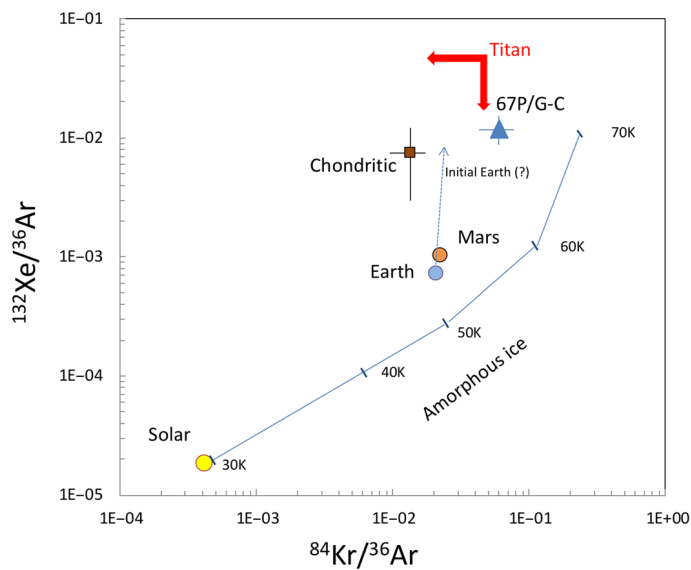

**Fig. 4. Noble gas relative abundances compared to other solar system reservoirs.** Sources of data: 67P/C-G: production rate ratios with 1-σ errors (SEM and calibration uncertainties) derived from Table 1, $^{84}Kr/^{36}Ar$ = 0.058 ± 0.013, and $^{132}Xe/^{36}Ar$ = 0.013 ± 0.003; solar: (18); Earth and Mars: (43); chondritic: CRPG compilation of CI and CM data; amorphous water ice: (24, 25). The blue arrow indicates the presumed composition of the initial atmosphere of Earth before secondary loss of xenon, and possibly krypton, to space through geological periods of time. The two red arrows represent upper limits for the $^{132}Xe/^{36}Ar$ and $^{84}Kr/^{36}Ar$ ratios measured in Titan by the Cassini-Huygens probe (23).

Similar to amorphous ice studies, the trapping and desorption capabilities of other ices like clathrate hydrates or polycrystalline ice must also be investigated in depth using appropriate gaseous mixtures. Many laboratory experiments, including species such as argon, krypton, and xenon, have been made in the 100 to 273 K range (37), but none of them are close to astrophysical conditions and compositions. Argon, krypton, and xenon abundances in clathrates formed in the protosolar nebula in the vicinity of the giant planets were predicted close to the observed values (38). However, the gaseous mixture used in this simulation is dominated by CO, which is not consistent with 67P/C-G observations where CO was much less abundant compared to $H_2O$ and $CO_2$. In addition, pressure and temperature conditions were chosen to reflect average nebular conditions and might differ from the actual region where 67P/C-G formed. They also performed the same computation for a $CH_4$-dominated mixture, which yielded very different relative noble gas abundances. These experiments show that the results depend strongly on the major molecule(s) and on pressure and temperature. No trapping or release data for noble gases exist for polycrystalline ices. Hence, additional laboratory results that are based on appropriate mixtures are needed for a comparison with ROSINA's observations. This includes the investigation to determine whether simultaneous release of $CO_2$ and noble gases is compatible with clathrates. Another well-known issue with the growth of clathrates in low-pressure environments in the outer (trans-Neptunian) protosolar nebula is their very slow formation kinetics, which can take longer than the disk lifetime, and is mainly due to the lack of available "fresh" ice (38). A way to overcome this issue is to assume that collisions between ice grains during planetesimal formation produce fresh ice at their surface and ease the clathrate formation process.

Finally, the starting composition of the gas might not have been solar, as suggested by xenon isotopic measurements (12), and the comparison with amorphous ice data, which assume a solar-like initial gas, may not be fully relevant. The same has to be considered in future experiments using also other types of ice. We think that the problem is far from being solved and that the present measurement of cometary noble gases, which has never been done before, will motivate further experimental and theoretical studies.

The Earth and Mars data points in Fig. 4 are off mixing lines that join solar, chondritic, or cometary compositions. These trends were taken as evidence for a cometary contribution to Earth and Mars (39) based on amorphous ice measurements (25). Recently, however, the xenon depletion of the terrestrial atmosphere has been attributed to preferential loss of this gas to space during geological periods of time (40, 41) based on the observation that the isotopic composition of xenon in the Archean Eon was intermediate between the primordial and modern atmosphere's one (the modern terrestrial atmospheric xenon being isotopically fractionated relative to primordial xenon). A more limited fraction of krypton might also have been lost (modern atmospheric krypton is also fractionated but to a lesser extent than xenon). A similar process might have also acted on Mars as martian atmospheric xenon is also depleted and isotopically fractionated. Thus, the initial composition of the terrestrial atmosphere (and possibly of the martian one) was presumably shifted toward higher Xe/Ar and Kr/Ar values ("initial Earth," blue arrow in Fig. 4), which would place the data points closer to a mixing line between chondritic and cometary. When corrected for xenon atmospheric loss on Earth, the noble gas composition of the "restored" Earth would fall between the average chondritic and the cometary compositions in Fig. 4. This location suggests that the atmosphere received contributions from both cosmochemical sources. This possibility is in full agreement with the independent evidence based on xenon isotopes that comets contributed 22 ± 5% cometary xenon to a 78 ± 5% chondritic xenon atmosphere (12).

Figure 4 shows the upper limits of possible Kr/Ar and Xe/Ar ratios estimated in Titan's atmosphere from measurements by the gas chromatograph–mass spectrometer (GC-MS) aboard the Cassini-Huygens probe. Among the primordial noble gases, only $^{36}$Ar has been firmly detected by the GC-MS with an atmospheric mole fraction of $\sim 2.06 \times 10^{-7}$ (23). On the other hand, only upper limits of $1 \times 10^{-8}$ have been inferred for the atmospheric mole fractions of krypton and xenon (we assume that the reported upper limits correspond to the major isotopes $^{84}$Kr and $^{132}$Xe). Some of the noble gases could have been trapped in the forming aerosols in Titan's atmosphere, which settled on the surface, thus lowering the krypton and xenon below the detection limit (42). A comparison between the upper limits of Xe/Ar and Kr/Ar ratios in Titan and 67P/C-G shows that the $^{84}$Kr/$^{36}$Ar ratio measured in the comet $(5.85 \pm 1.33) \times 10^{-2}$ overlaps within error bars with the upper limit $(4.85 \times 10^{-2})$ measured in Titan. In addition, the upper limit of the $^{132}$Xe/$^{36}$Ar ratio derived in Titan does not exclude 67P/C-G's value. Both, Titan and 67P/C-G, formed beyond the snowline in the protosolar nebula, and our data do not fully exclude a common origin for the building blocks from which both bodies agglomerated.

## MATERIALS AND METHODS
### ROSINA DFMS instrument

Neutral gas that entered the ion source was ionized by 45-eV electrons and subsequently accelerated into the analyzer section of the sensor. After a 90° deflection in an electrostatic analyzer and a 60° deflection in a permanent magnet, only ions with a very specific mass/charge can make it to the microchannel panel (MCP) detector (2). Ions impacting on the MCP trigger an avalanche of electrons, which are then collected and measured on a Linear Electron Detector Array (LEDA). For proper operation, an initial charge was applied to the LEDA that generates an offset that has to be subsequently removed (3, 14). The MCP/LEDA detector consists of two parallel arrays, row A and row B, for redundancy. Both rows contain 512 pixels each, which allow obtaining one mass spectrum per row around a preset mass in one go without changing any of the voltages.

Any given mass/charge ratio requires a specific set of potentials to be applied, and a single measurement lasts approximately 30 s. After the voltages were set, which took roughly 10 s, the signal was accumulated for ~20 s. This process was then repeated mass after mass, and a nominal mass scan from mass/charge 13 to 100 u/e took roughly 45 min. Therefore, DFMS was operated in a dedicated noble gas mode in which only a subset of the whole mass range was measured (details shown in table S4). Given the sequential measurement of different masses, DFMS was set to adjust the detector gain voltage based on ion current on the detector. For weak signals, including all noble gases (argon, krypton, and xenon), the highest amplification was chosen automatically. However, $H_2O$, $CO_2$, and CO (governing the signal on the mass/charge 28 u/e channel with $N_2$) are more abundant, and therefore, the spectra were obtained at lower gain amplifications. The relative amplifications were calibrated in the laboratory, and corresponding factors (~2.63 amplification of the signal per gain step) were taken into account when computing relative abundances with respect to $H_2O$, $CO_2$, and $N_2$. Furthermore, we restricted our analysis of the relative abundances of $^{36}Ar$ to $N_2$ and $N_2$ to $CO_2$ to measurements that were obtained within a maximum difference of three gain steps.

Not all channels on the MCP receive the same signal intensity over time, which leads to nonuniform aging depending, to first order, on the accumulated lifetime charge of each channel. This was monitored at regular intervals by varying the voltages such that a water beam was moved across the whole detector. From the variation of the signal, the sensitivities of each channel can then be derived. For the isotopic measurements reported here, however, the individual isotopes were measured one after the other and fell more or less on the exact same pixels, which, to first order, removed any errors associated with the aging process of the individual pixels. Row B was more sensitive compared to row A, and we therefore focus on row B in this work.

For the data analysis, first, the LEDA offset due to the initially applied charge was removed. Afterward, the pixel gain due to the different aging of the individual MCP channels was corrected for. Furthermore, mass fractionation in the mass analyzer has to be corrected for: In DFMS, the heavier species require a lower energy to make it through the analyzer section. Hence, also the impact velocity and thus yield at the MCP detector are lower. The correction was obtained from the laboratory calibration campaign.

The in-flight gas calibration unit flown together with the flight model of DFMS contained gases of neon, $CO_2$, and xenon. Therefore, the different noble gases (neon, argon, krypton, and xenon), together with other cometary volatile species including $H_2O$, $CO_2$, and $N_2$, which are discussed in this paper, were calibrated using the flight spare instrument in the laboratory. DFMS instrument sensitivities relate the abundances of the corresponding species inside the ion source to the measured signal of the MCP/LEDA detector used here. They were derived through cross-calibration using a reference Granville-Phillips Series 370 Stabil-Ion gauge at different pressures of the compound in question. Details on the calibration and the values can be found in the NASA Planetary Data System (PDS) and ESA Planetary Science Archive (PSA) archives (see the "Data and materials availability" section under Acknowledgments).

### Statistical errors

The statistical count error of the average counts per spectrum (for example, $\Delta^{80}Kr_{counts}$) is obtained from the square root of the average counts per spectrum (for example, $\Delta^{80}Kr$) divided by the number of spectra, $\Delta^{80}Kr_{counts} = \sqrt{^{80}Kr/N_{spectra}}$. Then follows a systematic error of the offset subtraction based on the peak width in number of pixels and the uncertainty of the offset per pixel, for example, $\Delta^{80}Kr_{offset} = N_{pixel} \times \Delta offset$. In the special case of $^{80}Kr$ follows the estimated error of 60% due to the strong interference of $C^{34}S_2$, which is close in mass and cannot be fully resolved by DFMS, and the same is true for $H^{79}Br$ (15), $\Delta^{80}Kr_{fit} = 0.6 \times {}^{80}Kr$. Therefore, uncertainties for $^{80}Kr$ are too large for any meaningful contribution to our analysis. The total error for the average counts is obtained from the square root of the sum of the squared individual errors, $\Delta^{80}Kr = \sqrt{\Delta^{80}Kr_{counts}^2 + \Delta^{80}Kr_{offset}^2 + \Delta^{80}Kr_{fit}^2}$. Finally, the errors after normalization to $^{84}Kr$ are obtained by error propagation, for example

$$\Delta\left(\frac{^{80}Kr}{^{84}Kr}\right) = \sqrt{\left(\frac{1}{^{84}Kr}\right)^2 (\Delta^{84}Kr)^2 + \left(\frac{^{80}Kr}{(^{84}Kr)^2}\right)^2 (\Delta^{84}Kr)^2}$$

For argon, corresponding errors are obtained through error propagation, that is, for $^{36}Ar/^{38}Ar = r \pm \Delta r = 5.4 \pm 1.4$ (10) follows

$$^{36}Ar/Ar = r/(r+1) = 5.4/(5.4 + 1.0) = 0.844$$

and

$$\Delta(^{36}Ar/Ar) = \frac{\Delta r}{(r+1)^2} = 0.034$$

More details and an example for xenon can be found in the Supplementary Materials of (12). The errors of the relative abundances between the different noble gases and between the noble gases and $H_2O$, $CO_2$, or $N_2$ were computed accordingly.

### Mixing model

We modeled the isotopic compositions of 67P/C-G krypton and xenon as the result of mixing between an exotic component rich in s-process isotopes and labeled here as X with a normal component resembling solar noble gases, for example, the N-component identified in presolar material [see (20, 22) and the Discussion section for justification and (11) for summary of data]. For one isotope $j$ of xenon, $^jXe$, the mass balance equation is

$$^jXe_O = {}^jXe_X + {}^jXe_N \quad (1)$$

where suffixes O, X, and N refer to cometary, exotic, and normal, respectively. Let $\alpha$ be the fraction of exotic $^jXe_X$ in cometary $^jXe_O$

$$\alpha = \frac{^jXe_X}{^jXe_O} \quad (2)$$

We also have

$$\frac{^jXe_N}{^jXe_X} = \frac{1 - \alpha}{\alpha} \quad (3)$$

Let us consider another isotope $j$ of xenon, the isotopic ratio $^iR = {^i\mathrm{Xe}}/{^j\mathrm{Xe}}$ is

$$^iR_O = \alpha\, ^iR_X + (1-\alpha)\, ^iR_N \quad (4)$$

with $\alpha$ defined in Eq. 2 as the fraction of the denominator isotope (here $^{132}$Xe) in X relative to O. A similar equation can be written for krypton

$$^j\mathrm{Kr}_O = {^j\mathrm{Kr}_X} + {^j\mathrm{Kr}_N} \quad (5)$$

$$\beta = \frac{^j\mathrm{Kr}_X}{^j\mathrm{Kr}_O} \quad (6)$$

$$\frac{^j\mathrm{Kr}_N}{^j\mathrm{Kr}_X} = \frac{1-\beta}{\beta} \quad (7)$$

Let us consider the Xe/Kr ratio. The mixing equation can be written as

$$\left(\frac{\mathrm{Xe}}{\mathrm{Kr}}\right)_O = \frac{\left(\frac{\mathrm{Xe}}{\mathrm{Kr}}\right)_X + \left(\frac{\mathrm{Xe}}{\mathrm{Kr}}\right)_N \left(\frac{\mathrm{Kr}_N}{\mathrm{Kr}_X}\right)}{1 + \left(\frac{\mathrm{Kr}_N}{\mathrm{Kr}_X}\right)} \quad (8)$$

which can be rewritten as

$$\left(\frac{\mathrm{Xe}}{\mathrm{Kr}}\right)_X = \frac{\left(\frac{\mathrm{Xe}}{\mathrm{Kr}}\right)_O - (1-\beta)\left(\frac{\mathrm{Xe}}{\mathrm{Kr}}\right)_N}{\beta} \quad (9)$$

This equation gives the Xe/Kr ratio of the exotic component required to yield the Xe/Kr ratio of the comet. In the following, the Xe/Kr ratios were normalized to the solar value. The normalized cometary ratio is 5, (main text) and let us assume that the normal component has a solar-like ratio [$(\mathrm{Xe}/\mathrm{Kr})_N = 1$]. For krypton, the best isotopic fit was obtained for $\beta = 0.05$ (main text and Fig. 1). It comes out that the exotic component should have a $(\mathrm{Xe}/\mathrm{Kr})_X$ ratio of 81 to fit the mixing equation, very rich indeed in xenon relative to krypton and solar.

We can also compute the fraction of $\mathrm{Xe}_N$ to cometary $\mathrm{Xe}_O$ for a $(\mathrm{Xe}/\mathrm{Kr})_X$ ratio of 81. Eq. 9 can be rewritten as

$$\left(\frac{\mathrm{Kr}}{\mathrm{Xe}}\right)_O = \frac{\left(\frac{\mathrm{Kr}}{\mathrm{Xe}}\right)_X + \left(\frac{\mathrm{Kr}}{\mathrm{Xe}}\right)_N \left(\frac{\mathrm{Xe}_N}{\mathrm{Xe}_X}\right)}{1 + \left(\frac{\mathrm{Xe}_N}{\mathrm{Xe}_X}\right)} \quad (10)$$

which can be rearranged as

$$\left(\frac{\mathrm{Xe}_N}{\mathrm{Xe}_X}\right) = \frac{\left(\frac{\mathrm{Kr}}{\mathrm{Xe}}\right)_X - \left(\frac{\mathrm{Kr}}{\mathrm{Xe}}\right)_O}{\left(\frac{\mathrm{Kr}}{\mathrm{Xe}}\right)_O - \left(\frac{\mathrm{Kr}}{\mathrm{Xe}}\right)_N} \quad (11)$$

which gives 0.234. The fraction of exotic xenon in cometary xenon, defined as $\alpha$ in Eq. 2 and computed using Eq. 3, is 0.81, which is identical to the fraction of 0.8 obtained to fit best the xenon isotopic data (fig. S5). Thus, two different approaches yield consistent results, giving strength to the mixing model developed in this contribution.

## SUPPLEMENTARY MATERIALS

Supplementary material for this article is available at http://advances.sciencemag.org/cgi/content/full/4/7/eaar6297/DC1

Fig. S1. Averaged ROSINA DFMS mass spectra containing the major krypton isotopes.
Fig. S2. Observation geometry for the noble gas observations in May 2016.
Fig. S3. Observation geometry before perihelion in May 2015 and early June 2015 used to derive the $N_2/H_2O$ ratio.
Fig. S4. $N_2$ versus $H_2O$ abundances for before perihelion in May 2015 and early June 2015.
Fig. S5. Mix of exotic X-Xe with normal N-Xe compared to 67P/C-G xenon.
Table S1. Krypton average counts per spectrum (~20-s integration time) including errors and below the associated isotopic ratios with respect to $^{84}$Kr with errors for the three periods.
Table S2. Ratio of $^{36}$Ar to total argon, $^{84}$Kr to total krypton, and $^{132}$Xe to total xenon.
Table S3. Bulk abundances of noble gases in the coma of comet 67P/C-G.
Table S4. Measured mass/charge ratios in the dedicated noble gas mode.

**Acknowledgments:** ROSINA would not give such outstanding results without the work of the many engineers, technicians, and scientists involved in the mission, in the Rosetta spacecraft, and in the ROSINA instrument team over the last 20 years whose contributions are acknowledged. These measurements could only be performed thanks to the skillful maneuvering of the spacecraft close to the comet by the operations team in Darmstadt. Rosetta is an ESA mission with contributions from its member states and NASA. We acknowledge herewith the work of the whole ESA Rosetta team. **Funding:** The following institutions and agencies supported this work: University of Bern was funded by the State of Bern, the Swiss National Science Foundation (200021_165869), and by the ESA PROgramme de Développement d'Expériences scientifiques (PRODEX) Program. Work at Max Planck Institute for Solar System Research was funded by the Max-Planck Society and Bundesministerium für Wirtschaft und Energie (contract 50QP1302) at Southwest Research Institute by Jet Propulsion Laboratory (JPL) (subcontract no. 1496541 and JPL subcontract to J.H.W. NAS703001TONMO710889) at the University of Michigan by NASA (contract JPL-1266313). This work was supported through the A*MIDEX project from the French National Research Agency [Agence Nationale de la Recherche (ANR)] (no. ANR-11-IDEX- 0001-02); by CNES (Centre National d'Etudes Spatiales) grants at IRAP (Institut de Recherche en Astrophysique et Planétologie), LATMOS (Laboratoire Atmosphères, Milieux, Observations Spatiales), LPC2E (Laboratoire de Physique et Chimie de l'Environnement et de l'Espace), LAM (Laboratoire d'Astrophysique de Marseille), and CRPG; and by the European Research Council (grant nos. 267255 and 695618 to B.M.) at Koninklijk Belgisch Instituut voor Ruimte-Aeronomie–Institut royal d'Aéronomie Spatiale de Belgique by the Belgian Science Policy Office via PRODEX/ROSINA PEA 90020. D.L. was supported by the Israel Ministry of Science and Technology through the Israel Space Agency (grant no. 3-11480). **Author contributions:** M.R. and K.A. performed data reduction. M.R., B.M., and K.A. analyzed the data. B.M., M.R., K.A., and O.M. interpreted the data and wrote the paper. K.A., M.R., B.M., and H.B. contributed to data interpretation and to the Supplementary Materials. K.A., H.B., A.B.-N., J.-J.B., C.B., U.C., M.C., J.D.K., B.F., S.A.F., S.G., T.I.G., K.C.H., E.K., A.K., D.L., L.L.R., U.M., B.M., O.M., T.O., H.R., M.R., T.S., C.-Y.T., J.H.W., and P.W. contributed to the ROSINA instrument, commented on and revised the manuscript. **Competing interests:** The authors declare that they have no competing interests. **Data and materials availability:** All ROSINA data and calibrated sensitivities used in this work have been released to the public Planetary Science Archive of ESA (www.cosmos.esa.int/web/psa/rosetta) and to the PDS archive of NASA. All data needed to evaluate the conclusions in the paper are present in the paper and/or the Supplementary Materials. Additional data related to this paper may be requested from the authors.

Submitted 6 December 2017
Accepted 24 May 2018
Published 4 July 2018
10.1126/sciadv.aar6297

**Citation:** M. Rubin, K. Altwegg, H. Balsiger, A. Bar-Nun, J.-J. Berthelier, C. Briois, U. Calmonte, M. Combi, J. De Keyser, B. Fiethe, S. A. Fuselier, S. Gasc, T. I. Gombosi, K. C. Hansen, E. Kopp, A. Korth, D. Laufer, L. Le Roy, U. Mall, B. Marty, O. Mousis, T. Owen, H. Rème, T. Sémon, C.-Y. Tzou, J. H. Waite, P. Wurz, Krypton isotopes and noble gas abundances in the coma of comet 67P/Churyumov-Gerasimenko. *Sci. Adv.* **4**, eaar6297 (2018).

**This PDF file includes:**

Fig. S1. Averaged ROSINA DFMS mass spectra containing the major krypton isotopes.
Fig. S2. Observation geometry for the noble gas observations in May 2016.
Fig. S3. Observation geometry before perihelion in May 2015 and early June 2015 used to derive the $N_2/H_2O$ ratio.
Fig. S4. $N_2$ versus $H_2O$ abundances for before perihelion in May 2015 and early June 2015.
Fig. S5. Mix of exotic X-Xe with normal N-Xe compared to 67P/C-G xenon.
Table S1. Krypton average counts per spectrum (~20-s integration time) including errors and below the associated isotopic ratios with respect to $^{84}Kr$ with errors for the three periods.
Table S2. Ratio of $^{36}Ar$ to total argon, $^{84}Kr$ to total krypton, and $^{132}Xe$ to total xenon.
Table S3. Bulk abundances of noble gases in the coma of comet 67P/C-G.
Table S4. Measured mass/charge ratios in the dedicated noble gas mode.

**H2: Supplementary Materials**

**Figures and Tables**

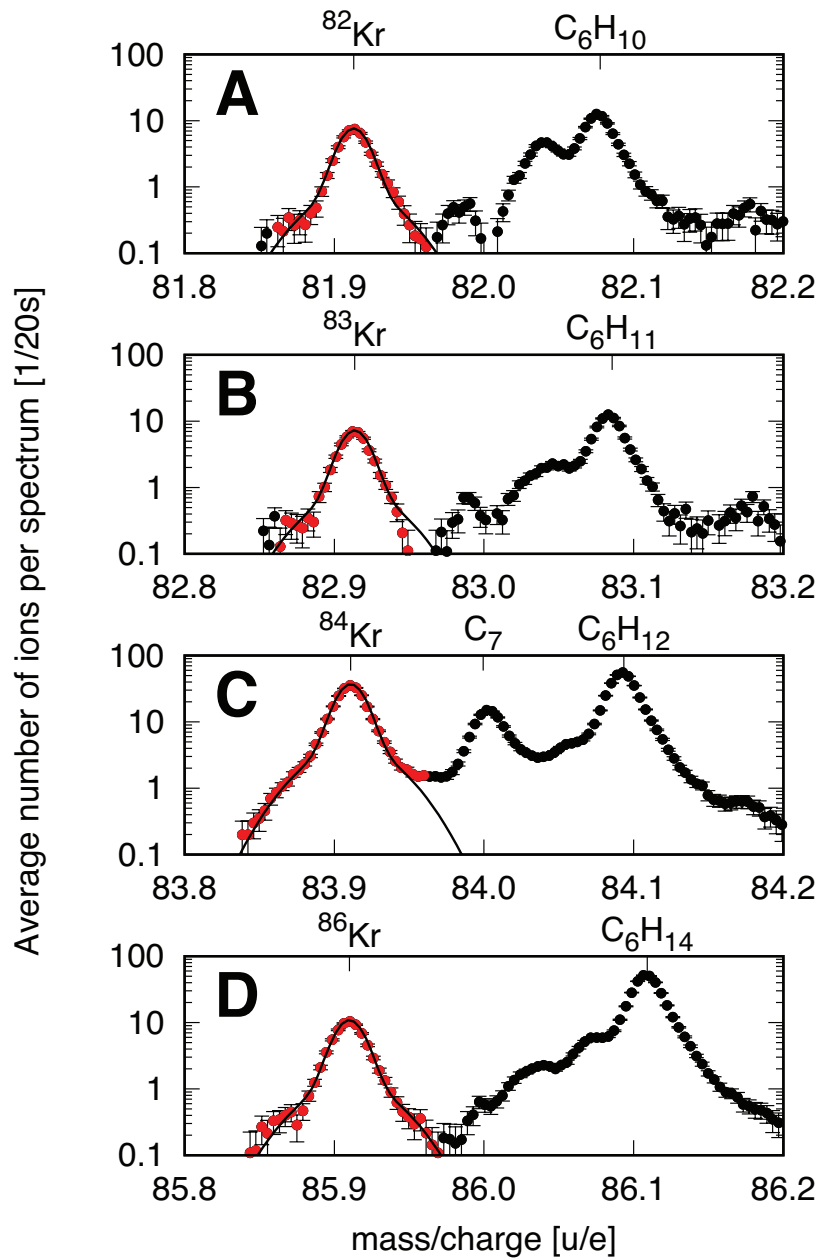

**Fig. S1. Averaged ROSINA DFMS mass spectra containing the major krypton isotopes.** $^{82}$Kr, $^{83}$Kr, $^{84}$Kr, and $^{86}$Kr are shown in panels A – D. Shown are the number of ions per spectra obtained from ~500 averaged measurements from 14 to 31 May 2016. The sum of the red measurement points corresponds to the total signal of the corresponding Kr signal reported here. The solid black line shows the fitted peak shape based on the $^{84}$Kr peak and is for visual aid only. The errors represent $\pm\sqrt{}$ due to count statistics.

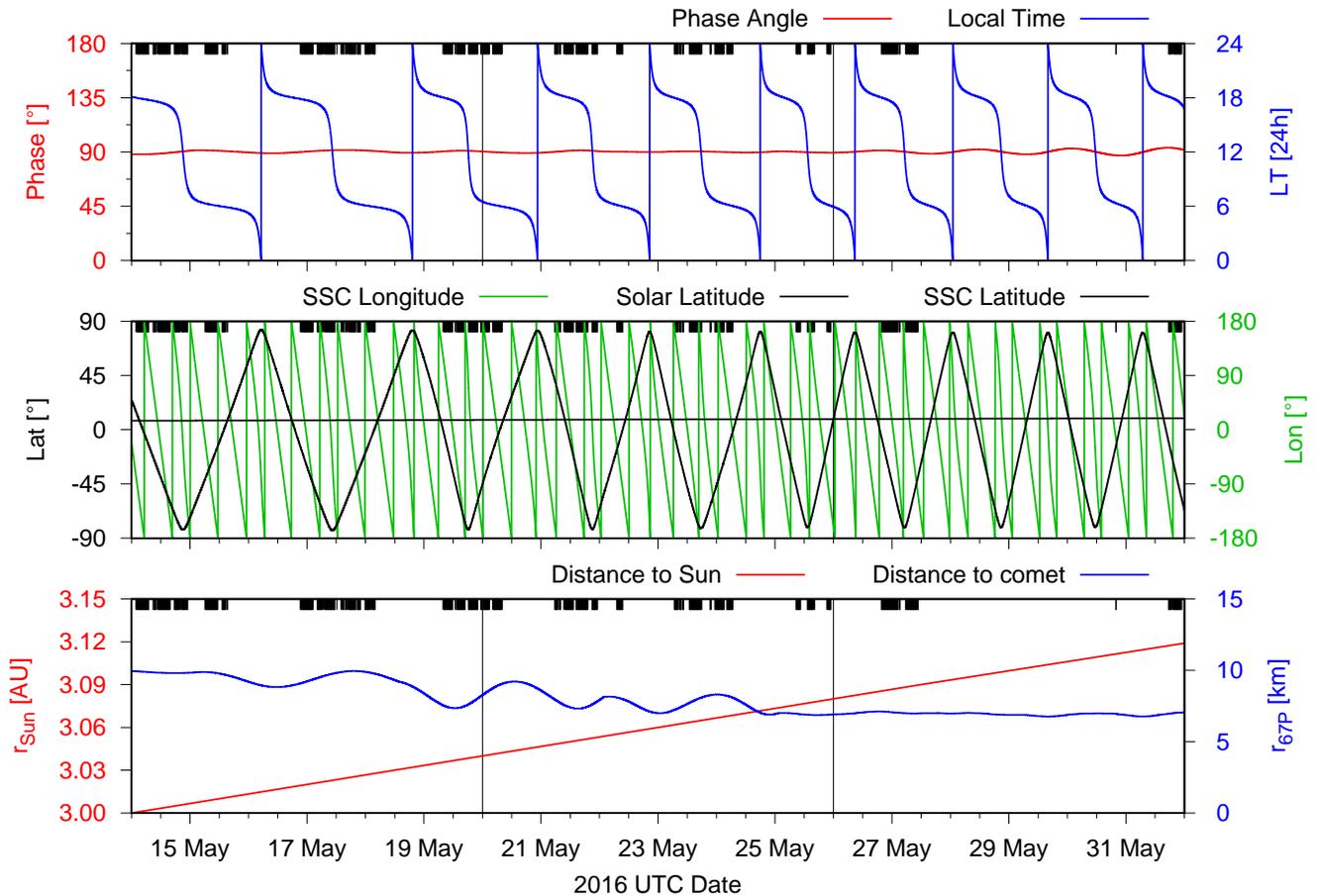

**Fig. S2. Observation geometry for the noble gas observations in May 2016.** The top panel shows the phase angle of Rosetta and the local time at the sub spacecraft point. The middle panel shows the longitude and latitude of the sub spacecraft point and the latitude of the sub solar point. The bottom panel shows the heliocentric and cometocentric distances. Black vertical lines indicate the start date of the three overlapping periods analyzed in this work: (i) 14 – 31 May, (ii) 20 – 31 May, and (iii) 26 – 31 May. The black tick marks on top of each panel correspond to one noble gas mode each (cf. table S4)

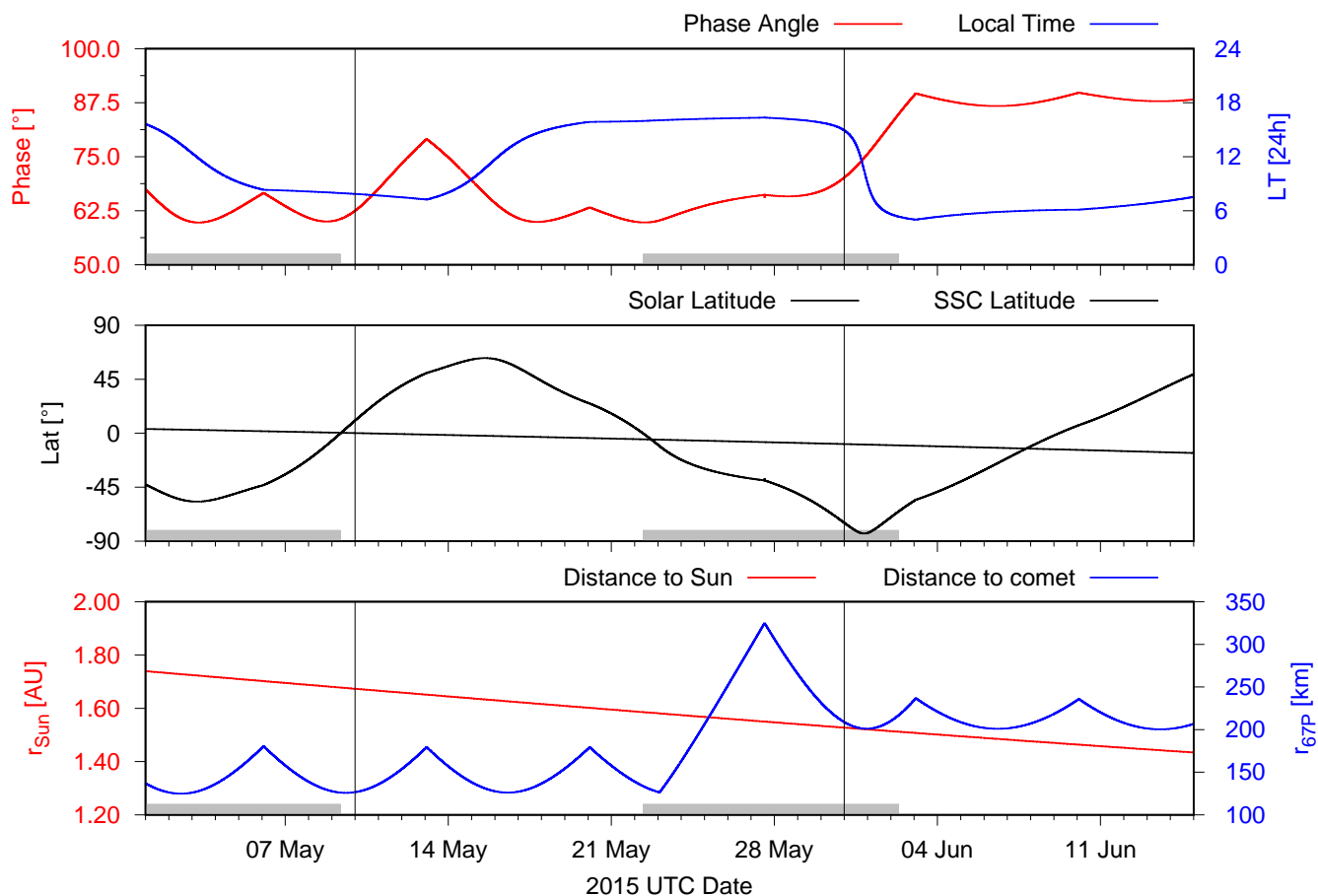

**Fig. S3. Observation geometry before perihelion in May 2015 and early June 2015 used to derive the $N_2/H_2O$ ratio.** The top panel shows the phase angle of Rosetta and the local time at the sub spacecraft point. The middle panel shows the longitude and latitude of the sub spacecraft point and the latitude of the sub solar point. The bottom panel shows the heliocentric and cometocentric distances. The grey bars at the bottom of each panel indicate times when Rosetta is on the dayside of the comet (phase angle < 85°) and above the at the time more active southern hemisphere (latitude < 0°). The black vertical lines indicate the period of 10 – 30 May identified as suitable to derive bulk abundances.

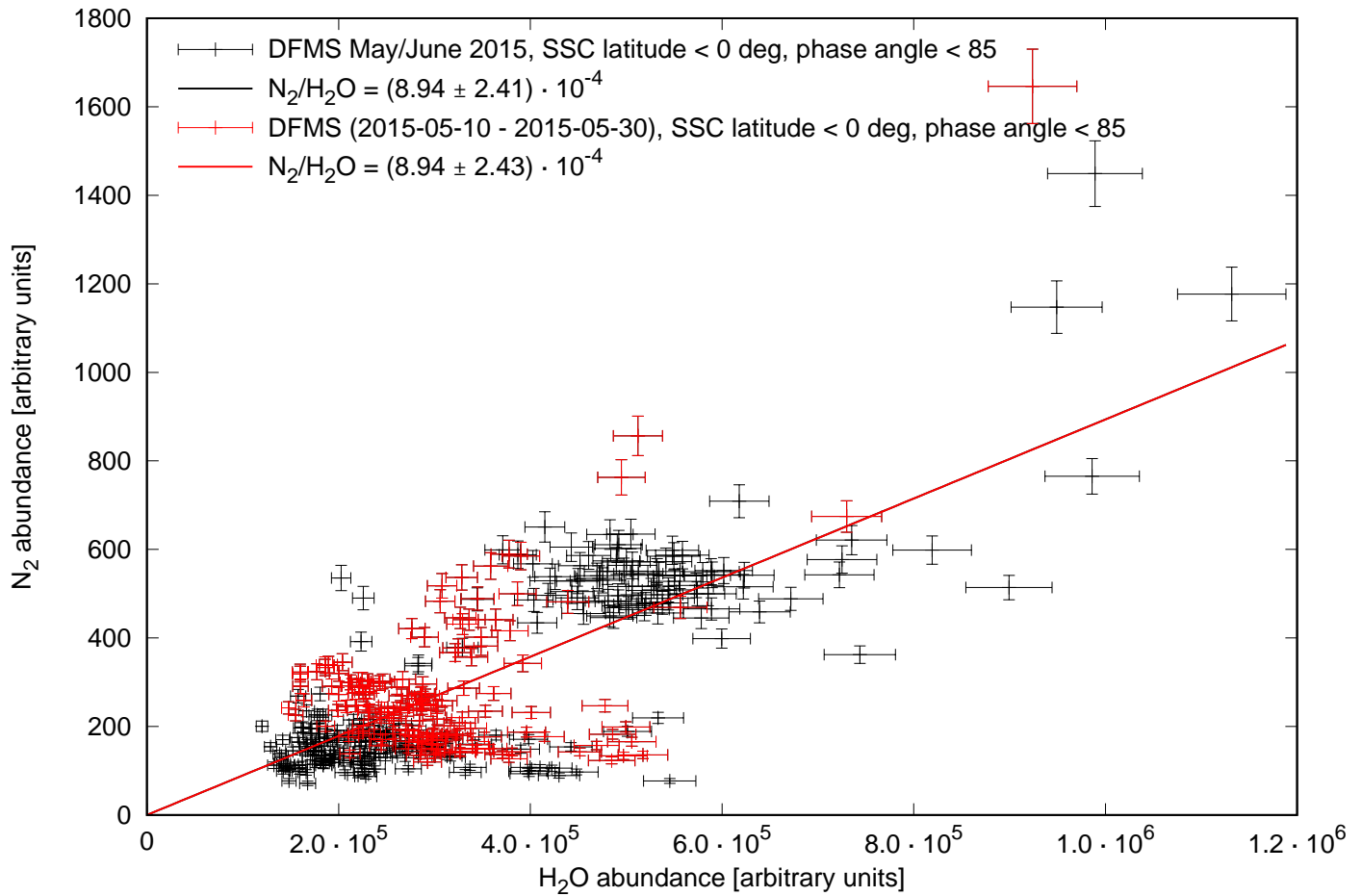

**Fig. S4. $N_2$ versus $H_2O$ abundances for before perihelion in May 2015 and early June 2015.**
The red points denote subset for the period of 10 – 30 May identified as a suitable period to derive bulk abundances (cf. fig. S3). Only measurements on the dayside of the comet (phase angle < 85°) and above the southern active hemisphere (latitude < 0°) have been taken into account. Linear fits using a least-squares approach with uniform weights have been applied to both datasets and yield very similar slopes for the bulk abundance of $N_2$ with respect to $H_2O$. The errors in the data points include 1-σ statistical errors. For the slopes additionally 18% calibration uncertainty (sensitivity and fragmentation pattern) and 20% gain error have been included.

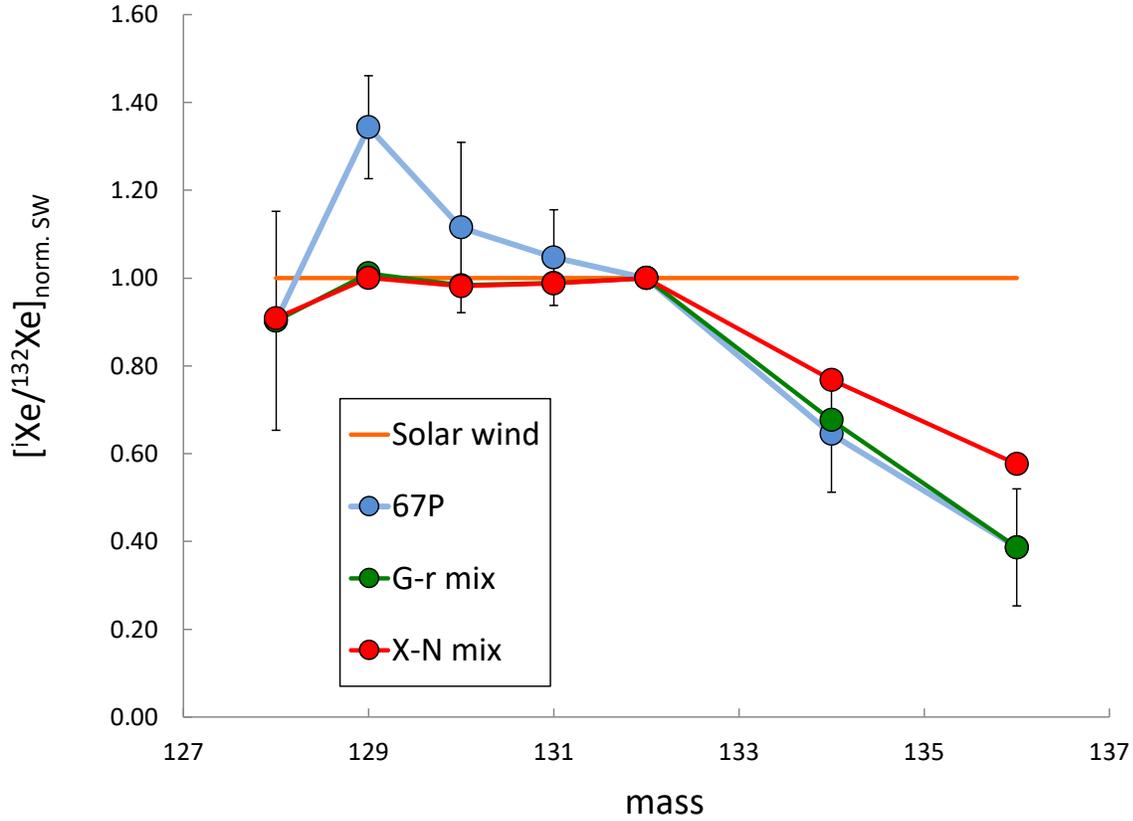

**Fig. S5. Mix of exotic X-Xe with normal N-Xe compared to 67P/C-G xenon.** We use an N process composition (*20*) and the X component is defined as a mixture of s-process Xe (here taken as G-Xe) with r-process Xe (*11*). In our earlier work on xenon (*11*) the s-process composition was taken from (*19*) and here we take the G-Xe composition (*10*), which is close to the former, for the sake of consistence with the case of krypton. The best fit is obtained for a proportion of 80% X-Xe in cometary xenon. The proportion of the mixture are similar to those proposed for xenon (*11*), however, one should consider a X composition more depleted in $^{134}$Xe and $^{136}$Xe to give cometary Xe when adding N-Xe (which is not depleted in these isotopes). Nevertheless, a satisfactory mix within errors is obtained as the proportion of N-Xe is low.

**Table S1. Krypton average counts per spectrum (~20-s integration time) including errors and below the associated isotopic ratios with respect to $^{84}$Kr with errors for the three periods.** The results in this paper are based on the 14 – 31 May 2016 period (underlined). Errors reflect 1-σ s.e.m. and calibration uncertainties.

|  | $^{80}$Kr | $^{82}$Kr | $^{83}$Kr | $^{84}$Kr | $^{86}$Kr |
|---|---|---|---|---|---|
| **14 – 31 May** | | | | | |
| counts | 9.387 | 53.399 | 48.353 | 258.652 | 74.322 |
| count error | 0.136 | 0.323 | 0.308 | 0.714 | 0.383 |
| offset error | 1.050 | 2.900 | 2.700 | 3.400 | 3.400 |
| fit error | 5.632 | 0.000 | 0.000 | 0.000 | 0.000 |
| total error | 5.731 | 2.918 | 2.717 | 3.474 | 3.422 |
| **20 – 31 May** | | | | | |
| counts | 12.895 | 60.932 | 56.336 | 297.822 | 87.407 |
| count error | 0.223 | 0.480 | 0.465 | 1.068 | 0.583 |
| offset error | 1.050 | 4.650 | 4.050 | 4.950 | 5.100 |
| fit error | 7.737 | 0.000 | 0.000 | 0.000 | 0.000 |
| total error | 6.536 | 4.675 | 4.077 | 5.064 | 5.133 |
| **26 – 31 May** | | | | | |
| counts | 15.222 | 82.140 | 76.843 | 414.043 | 118.580 |
| count error | 0.457 | 1.047 | 1.026 | 2.382 | 1.283 |
| offset error | 1.050 | 5.800 | 5.000 | 6.800 | 6.400 |
| fit error | 9.133 | 0.000 | 0.000 | 0.000 | 0.000 |
| total error | 6.196 | 5.894 | 5.104 | 7.205 | 6.527 |
|  | $^{80}$Kr/$^{84}$Kr | $^{82}$Kr/$^{84}$Kr | $^{83}$Kr/$^{84}$Kr | $^{84}$Kr/$^{84}$Kr | $^{86}$Kr/$^{84}$Kr |
| **14 – 31 May** | | | | | |
| ratio | <u>0.036</u> | <u>0.206</u> | <u>0.187</u> | <u>1.0</u> | <u>0.287</u> |
| ratio error | <u>0.022</u> | <u>0.012</u> | <u>0.011</u> | | <u>0.014</u> |
| **20 – 31 May** | | | | | |
| ratio | 0.043 | 0.205 | 0.189 | 1.0 | 0.293 |
| ratio error | 0.026 | 0.016 | 0.014 | | 0.018 |
| **26 – 31 May** | | | | | |
| ratio | 0.037 | 0.198 | 0.186 | 1.0 | 0.286 |
| ratio error | 0.022 | 0.015 | 0.013 | | 0.017 |

**Table S2. Ratio of $^{36}$Ar to total argon, $^{84}$Kr to total krypton, and $^{132}$Xe to total xenon.** Kr and Xe results were based on the 14 – 31 May 2016 period. Errors contain 1-σ s.e.m. and calibration uncertainties.

| Isotope | Ratio | Reference |
|---|---|---|
| $^{36}\text{Ar}/(^{36}\text{Ar} + {}^{38}\text{Ar})$ | $0.844 \pm 0.034$ | (13) |
| $^{84}\text{Kr}/\sum_i {}^i\text{Kr}$ | $0.582 \pm 0.010$ | this work |
| $^{132}\text{Xe}/\sum_i {}^i\text{Xe}$ | $0.253 \pm 0.012$ | (11) |

**Table S3. Bulk abundances of noble gases in the coma of comet 67P/C-G.** The second column gives the density ratios measured at the location of Rosetta, which is equivalent to the comet's production rate assuming equal outgassing velocities applicable in a collisional coma. The last column assumes a collisionless coma and the ratio $X_1/X_2$ is multiplied by $\sqrt{m_2/m_1}$ to account for different outgassing velocities based on a thermal expansion approximation. $N_2/H_2O$ has been derived pre-perihelion in May/June 2015. $^{36}Ar/N_2$ was obtained in May 2016 at 6 km from the center of the comet. Ar, Kr, and Xe with respect to $H_2O$ have been derived from $N_2/H_2O$, $^{36}Ar/N_2$, table S2, and Table 1. Errors reflect 1-σ s.e.m. and calibration uncertainties.

| Species | Density ratio at Rosetta | $\sqrt{m_2/m_1}$ | Production rate ratio 67P |
|---|---|---|---|
| $N_2/H_2O$ | $(8.9 \pm 2.4) \cdot 10^{-4}$ | 0.802 | $(7.2 \pm 2.0) \cdot 10^{-4}$ |
| $^{36}Ar/N_2$ | $(6.2 \pm 1.7) \cdot 10^{-3}$ | 0.882 | $(5.5 \pm 1.5) \cdot 10^{-3}$ |
| $^{36}Ar/H_2O$ | $(5.5 \pm 2.1) \cdot 10^{-6}$ | 0.707 | $(3.9 \pm 1.5) \cdot 10^{-6}$ |
| $Ar/H_2O$ | $(6.6 \pm 2.5) \cdot 10^{-6}$ | 0.707 | $(4.6 \pm 1.8) \cdot 10^{-6}$ |
| $Kr/H_2O$ | $(8.5 \pm 3.8) \cdot 10^{-7}$ | 0.463 | $(3.9 \pm 1.8) \cdot 10^{-7}$ |
| $Xe/H_2O$ | $(5.3 \pm 2.4) \cdot 10^{-7}$ | 0.369 | $(1.9 \pm 0.9) \cdot 10^{-7}$ |

**Table S4. Measured mass/charge ratios in the dedicated noble gas mode.** Mode duration ~14 min.

| Mass/charge [u/e] | Justification |
|---|---|
| 18 | Water for reference |
| 20, 22 | Neon isotopes |
| 36, 38 | Argon isotopes |
| 40, 41, 42, 43 | Double-charged krypton isotopes |
| 64, 65, 66, 68 | Double-charged xenon isotopes |
| 78, 80, 82, 83, 84, 86 | Krypton isotopes |
| 124, 126, 128, 129, 130, 131, 132, 134, 136 | Xenon isotopes |